# Overcoming thermal noise in non-volatile spin wave logic

Sourav Dutta[*,1], Dmitri E. Nikonov[2], Sasikanth Manipatruni[2], Ian A. Young[2], and Azad Naeemi[1]

[1]School of Electrical and Computer Engineering, Georgia Institute of Technology, Atlanta, GA 30332 USA

[2]Components Research, Intel Corporation, Hillsboro, OR 97124 USA

*Correspondence to sdutta38@gatech.edu

**Spin waves are propagating disturbances in magnetically ordered materials, analogous to lattice waves in solid systems and are often described from a quasiparticle point of view as magnons. The attractive advantages of Joule-heat-free transmission of information, utilization of the phase of the wave as an additional degree of freedom and lower footprint area compared to conventional charge-based devices have made spin waves or magnon spintronics a promising candidate for beyond-CMOS wave-based computation. However, any practical realization of an all-magnon based computing system must undergo the essential steps of a careful selection of materials and demonstrate robustness with respect to thermal noise or variability. Here, we aim at identifying suitable materials and theoretically demonstrate the possibility of achieving error-free clocked non-volatile spin wave logic device, even in the presence of thermal noise and clock jitter or clock skew.**

In recent years, information processing circuits based on spin waves have been the subject of intense research as they hold promise to augment complementary metal oxide semiconductor (CMOS) circuits and to open up a new horizon in extending Moore's law well into the future [1-3]. Spin waves, a collective oscillation of electron spins in a ferromagnetic metal or insulator, allow charge-free transmission of information and a novel wave-based computing paradigm exploiting wave interference and nonlinear wave interactions [4-10]. The recent advances in voltage-controlled magnetoelectric (ME) devices, which can switch the magnetization with reduced energy dissipation compared to current-controlled devices, have provided an alternative pathway for excitation and detection of spin waves compared to inductive [5,11-13] or spin-torque [14,15] excitation. While voltage-driven strain-mediated spin wave generation has been experimentally demonstrated [16,17] and theoretically studied in [18,19], more research is needed to experimentally develop voltage driven spin wave generation and detection as a competitive technology. In addition, there are many requirements that any novel computing platform must meet before it can be adopted for use in real circuits and even before major investments in research and development become justifiable. While the requirements of gain, concatenability, feedback prevention and logic function completeness has been addressed in a recent work [19], it has now become imperative to explore the question of robustness with respect to thermal noise and variability. To bring these schemes to their practical realization, the identification of suitable materials that can enable the experimental implementation of the developed ideas assumes critical importance. Furthermore, one of the promising attributes of magnetic devices is their non-volatility allowing zero-static power dissipation [20-23] and enabling implementation of logic-in-memory architectures [24-26]. As such, it is highly desired that a promising spin wave circuit can readily take advantage of this inherent feature of magnets. In this work, we focus on identifying suitable materials and set forth a set of design rules to achieve a thermally reliable clocked non-volatile spin wave device that meets all the requirements for logic circuits.

## Results
### Building blocks
We start by first describing the basic building blocks for a spin wave logic device and identifying suitable materials based on experimental demonstrations. The major ingredients are (i) ME cell operating as a spin wave transmitter and detector with in-plane stable magnetization states, and (ii) a spin wave bus (SWB) having perpendicular magnetic anisotropy (PMA) that acts as a conduit for information transmission. The choice of mutually orthogonal spin configuration of SWB and ME cell stems from the requirements of non-volatility and non-reciprocity [19]. A comparison with alternative spin configuration of ME-SWB system is provided in supplementary section S1.

### Spin wave bus (SWB)
SWBs are usually fabricated in the form of in-plane magnetized narrow magnetic stripes with commonly used materials like permalloy ($Ni_{81}Fe_{19}$) [3] and yttrium-iron-garnet ($Y_3Fe_5O_{12}$) [27,28] that provide low gilbert damping. However, recent works on spin waves have highlighted the preference of using an out-of-plane magnetized over in-plane magnetized SWB owing to several advantages. Firstly, it is possible to overcome the limitations of broken translational symmetry and anisotropic dispersion relation of backward volume spin waves that give rise to scattering processes where the waves interfere [29]. Secondly, it is possible to locally control the internal magnetic field via application of voltage controlled magnetic anisotropy (VCMA) [30,31]. While several experimental works have used a few hundred milli-Tesla (mT) of magnetic field for out-of-plane biasing, the usage of such an external magnetic field is incompatible with integrated device and circuitry. Multilayers like [Co/Ni] are potential candidates that can offer a bias-free out-of-plane magnetic configuration because of their inherent interface anisotropy arising from the spin-orbit interaction at the interface [32-36] and sustain a propagating spin wave [37,38] via their low damping [39-42]. Note that we aim at obtaining a low PMA to minimize spin wave attenuation. Using a phenomenological treatment along with experimentally determined parameters [32-36], we calculate an effective PMA of 0.4 $MJ/m^3$ for a [Co(0.4 nm)/Ni(0.8 nm)]$_{10}$ multilayer (Fig. 1(a)). Detailed calculations are provided in section S2 of the supplementary information. We would like emphasis that the focus of this work is on the principle of robust switching of nanomagnets by spin waves and we have chosen Co/Ni multilayers only as a convenient and well-studied material example. Although we consider idealized multilayers in our simulations characterized by low Gilbert damping as has been experimentally reported [39-42], the extrinsic damping arising from sources like two-magnon scattering due to sample-inhomogeneity and spin pumping at the heavy metal/ferromagnet interface can give rise to additional spin wave damping. Heusler alloy ($Co_2Fe_{0.4}Mn_{0.6}Si$) exhibiting low magnetic damping [43,44] may provide an attractive alternative for low-loss spin wave channel.

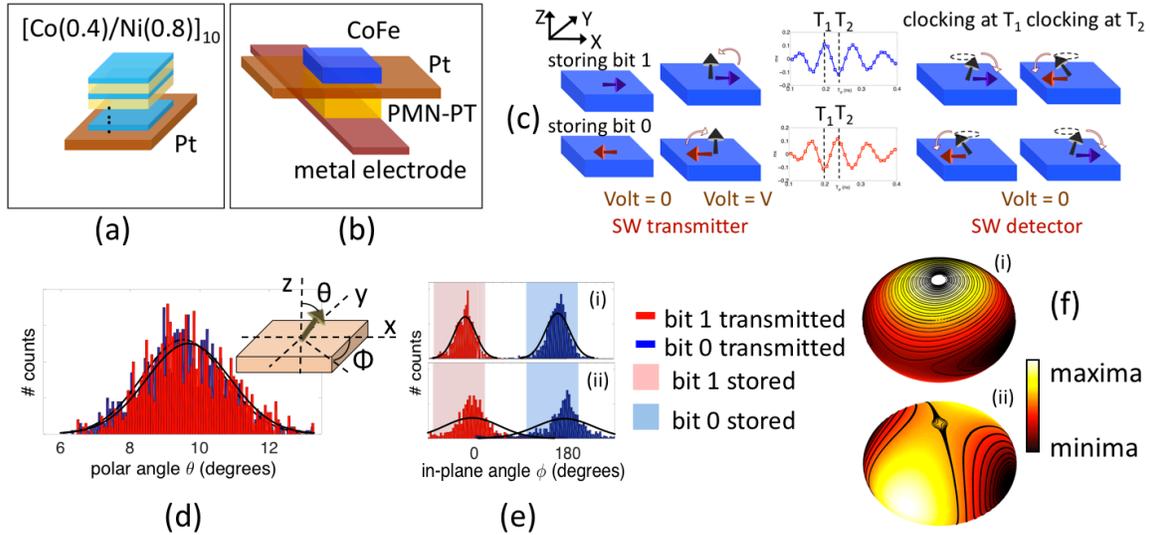

Figure 1: Basic building blocks for a spin wave logic device - (a) PMA $[Co(0.4)/Ni(0.8)]_{10}$ multilayer spin wave bus and (b) ME cell comprising of a magnetostrictive $Co_{60}Fe_{40}$ layer grown on (001)PMN-PT ferroelectric layer. (c) Working principle of the spin wave device, based on voltage-controlled strain-mediated magnetization switching, (d),(e) Gaussian distribution of the amplitude ($\theta = \cos^{-1} m_z$) and phase ($\phi = \tan^{-1}(m_y/m_x)$) of the arriving spin wave, detected at the falling edge of the clock, (f) energy landscape and magnetization relaxation trajectory from out-of-plane to in-plane energy minima for the case of (i) a nanomagnet with in-plane stable magnetization states, and (ii) lowered out-of-plane energy barrier resulting in the interchange of the position of the saddle point and the energy maxima.

**Magneto-electric (ME) cell**

The most fundamental computing block of a magnonic logic is the magneto-electric (ME) cell that acts as a spin wave transmitter, detector and also serves as a non-volatile memory element [19]. As shown in Fig. 1(b), it is a heterostructure consisting of a ferroelectric or piezoelectric material sandwiched between two metallic electrodes and a top magnetostrictive ferromagnetic layer. With ultra-low power dissipation being the ultimate goal in mind, the target piezoelectric material must possess a high piezoelectric coefficient ($d_{31}$) while the magnetic layer must display a high magnetostrictive coefficient ($\lambda$) simultaneously. Additionally, compatibility between the chosen materials must be ensured in order to implement the device experimentally. We explore a wide-range of theoretically and experimentally studied magnetostrictive (Ni, $CoFe_2O_4$, CoFeB, CoFe, CoFeV, Terfenol-D, FeGa, FeGaB, $Fe_3O_4$ and $NiFe_2O_4$) and ferroelectric (PMN-PT, PZN-PT, PZT, BaTiO3) materials and provide a comprehensive map of the magnetostrictive and piezoelectric coefficients including their compatibility. (Details provided in supplementary section S4). For the rest of the paper, we focus on $Co_{60}Fe_{40}$ that has been reported to display an enhanced magnetostriction at the (fcc+bcc)/bcc phase boundary [45] with effective $\lambda$ of 200 ppm, grown on (001) PMN-PT. The chosen combination allows one to reach a large product of coupling coefficient with added advantage of a much more mature fabrication process for CoFe compared with that of say, Terfenol-D.

**Device operation**
The working principle, based on voltage-controlled strain-mediated magnetization switching, is illustrated in Fig. 1(c). We designate a logic or bit "1" and "0" to the magnetization states pointing in the +x or -x direction, respectively. On application of a voltage, an in-plane to out-of-plane magnetization switching of the transmitter ME cell excites spin waves with the information encoded in the phase of the waves. The latter is then translated into the magnetization orientation of the detector ME cell via a phase-dependent deterministic switching as the voltage of the detector is switched off. From here on, we would refer to the time when the voltage of the detector ME cell is switched off as the "clocking time". We set the clocking time equal to the per stage propagation delay of the spin wave signal. Further details on the working principle and mathematical modeling are provided in supplementary sections S6, S7 and in [19,46]. Depending upon the time of clocking, we end up with the detector ME cell's magnetization assuming either of the two stable magnetization states. In other words, we can define the logic function of the SW device (buffer or inverter) simply by choosing the appropriate time of clocking. This scheme is in contrast to prior work on magnonic logic that relies on the length of the interconnect compared to the wavelength of the spin wave [6] and offers the possibility of having magnonic reconfigurable logic. The proposed device concept is universal in the sense that alternative mechanisms for 90° magnetization switching like VCMA can also be used instead of magnetostriction (see supplementary section S8 for comparison).

**Thermal reliability**
Thermal noise has constantly plagued the field of spintronics, influencing magnetic retention, read and write failures. The dynamic variability introduced by thermal fluctuations poses a serious threat to the performance of spintronics logic and memory. The effect of thermal noise on spin wave logic is twofold: (a) introduction of phase noise by randomizing the phase of the propagating spin wave, and (b) affecting the trajectory of the magnetization dynamics of the ME cell during the course of spin wave excitation and detection. The effect is seen to be most crucial during the course of detection. Firstly, the amplitude ($\theta = \cos^{-1} m_z$) and phase ($\phi = \tan^{-1}(m_y/m_x)$) of the arriving spin wave, detected at the falling edge of the clock displays a Gaussian distribution around the mean value expected without any thermal noise as shown in Figs. 1(d) and (e). Fig. 1(e,i) shows the case of an error-free logic function acting as a buffer where all detected phase ($\phi$) falls within the highlighted windows of deterministic switching. Hence, all the bit "1"s transmitted are stored as bit "1"s and same for bit "0"s. On the contrary, Fig. 1(e,ii) depicts a more erroneous case where the detected phase ($\phi$) spreads over both the windows giving rise to a situation where some of the transmitted bit "1"s get detected as a bit "0"s and so on. Also note that the white gap separating the regions of deterministic detection of bits "1"s and "0"s represents a non-deterministic situation as explained later.

Additionally, the out-of-plane to in-plane magnetization relaxation trajectory during the time of detection is sensitive to thermal fluctuations and a small variation can cause the switching to become non-deterministic. In a previous work, a scheme for achieving phase-dependent deterministic switching of the ME spin wave detector by modifying the energy landscape via compensation of demagnetization was introduced [46]. As shown in Fig. 1(f,i) the presence of an energy maxima in the out-of-plane +z direction causes the magnetization trajectory to be highly precessional following the constant energy orbits. If the energy landscape can be modified by lowering the out-of-plane energy barrier such that the position of the saddle point and the energy maxima gets interchanged, a more damped trajectory is obtained with the direction of the switching dependent solely on the initial angle, i.e., the phase of the spin wave (Fig. 1(f,ii)). Here, we investigate two viable options for translating the "theoretical idea" of phase-dependent switching of the spin wave detector to a "practical realization" of a thermally reliable magnonic device by - (a) using the built-in strain in the ME cell for compensation of the demagnetization, and (b) using an exchange-spring structure which inherently modifies the energy landscape of the ME cell magnet as desired.

**Built-in strain**
The first possible route is to take advantage of the "built-in strain". Fig. 2(a) illustrates one possible layout of a spin wave logic circuit, with the main building blocks - ME cell and PMA SWB highlighted in Fig. 2(b). Recent works on the growth and characterization of relaxor ferroelectric materials have demonstrated the possibility to engineer a desired built-in strain in a thin ferroelectric film grown epitaxially on an appropriate substrate. This misfit strain arising from the lattice mismatch and/or thermal expansion coefficient mismatch between the film and the substrate can be as high as -0.42% for (001) $Pb(Zr_{1-x}Ti_x)O_3$ (PZT) grown on $SrTiO_3$ (SRO) substrate [47] and -0.46% for (001) $0.9(Pb(Mg_{1/3}Nb_{2/3})O_3)$-$0.1[PbTiO_3]$(0.9PMN- 0.1PT) grown on (001)$LaAlO3$ substrate [48]. The degree of in-plane misfit strain can be varied by using different substrates like (La,Sr)(Al,Ta)O3(LSAT), SrTiO3(STO) and MgO. Here, we consider the scenario of an epitaxially grown (001) PMN-PT on an appropriate substrate capable of producing a small built-in strain of -0.31% to -0.37%. The PMN-Pt layer is sandwiched between a bottom metallic electrode and a top thin layer of Pt. The Pt is assumed to be thin enough to allow an efficient strain transfer to the top magnetostrictive ferromagnetic layer of $Co_{60}Fe_{40}$.

To better elucidate the impact of this built-in strain ($\epsilon_{res}$) on the switching error, we first look at the energy landscapes of the ME cell's magnetic layer under a zero $\epsilon_{res}$ condition and under $\epsilon_{res}$ = -0.35%. With no strain present, the magnetic layer has energy maxima in the out-of-plane axis (z) while the minima and the saddle points are along the x and y axis respectively. The presence of a small strain (less than the critical strain for PMA) manifests itself as a reduction of the out-of-plane energy barrier by introducing a small perpendicular anisotropy less than that of the shape anisotropy. As the out-of-plane energy barrier becomes less than the in-plane energy barrier, the positions of the energy maxima and the saddle points gets interchanged (Fig. 2c). This results in a change of the magnetization relaxation dynamics from a highly precessional one to a more damped trajectory, being more strongly dependent on the initial magnetization angles, or in other words, the phase of the arriving spin waves as highlighted earlier in Fig. 1(f). For $|\epsilon_{res}|$ > 0.4%, the energy barrier between the stable in-plane magnetization states is markedly reduced resulting in loss of non-volatility. Beyond the critical strain of -0.48%, the magnet becomes PMA (see supplementary section S9).

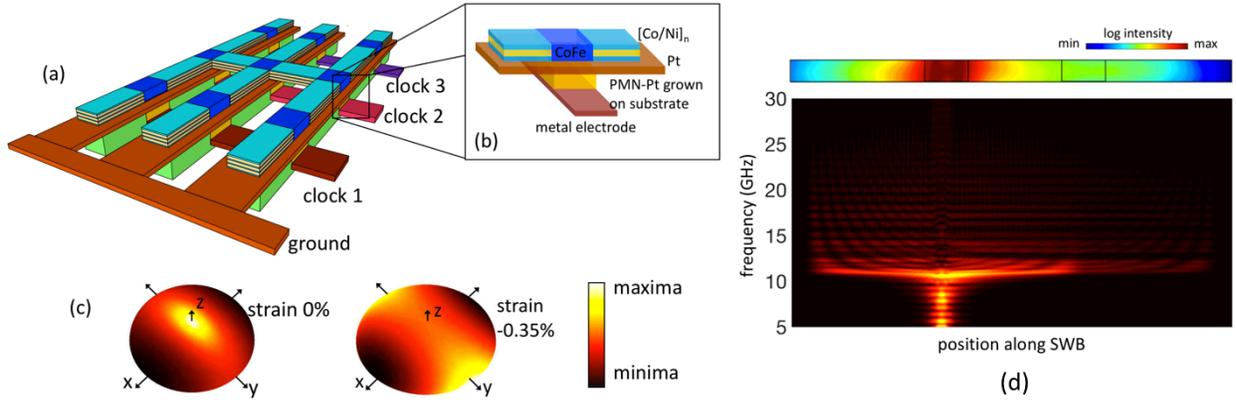

Figure 2: (a) Illustration of a possible layout of a spin wave logic circuit, (b) Details of the main building blocks - ME cell and PMA SWB. The ferroelectric PMN-PT is assumed to be epitaxially grown on an appropriate substrate in order to produce a small built-in strain. (c) Energy landscape of the CoFe layer of the ME cell under the case of 0 and -0.35% built-in strain. (d) Spin wave transmission from the transmitter ME cell through the SWB to the detector ME cell and the frequency spectra along the length of the SWB, obtained from FFT of the *x*-component of the magnetization

Note that there is butt-coupling of the [Co/Ni] multilayer waveguide and the CoFe layer of the ME cell. The voltage-driven strain-mediated magnetization switching of the transmitter ME cell from the in-plane to the out-of-plane configuration excites spin waves over a wide range of frequencies, as has been shown earlier in ref. [49]. However, only those frequencies which are above the cutoff frequency of the [Co/Ni] multilayer and CoFe layer are allowed to penetrate and propagate through the SWB and ME cell. The dispersion relation for the forward volume spin wave is calculated as $\omega^2 = \omega_0 \left[\omega_0 + \omega_M \left(1 - \frac{1-e^{-kd}}{kd}\right)\right]$ [50] where $d$ is the thickness, $\omega_0 = \gamma H_K = \gamma(H_{aniso} - H_{demag})$ involves the effective out-of-plane internal magnetic field and $\omega_M = \gamma M_S$ where $M_S$ is the saturation magnetization. The values of $M_S = 790 \, kA/m$ and $K_{PMA} = 0.4 \, MJ/m^3$ for our [Co/Ni] are very similar to the $M_S = 800 \, kA/m$ and voltage induced $K_{ME} = 0.3 - 0.42 \, MJ/m^3$ for CoFe. Hence, the minimum cut-off frequencies of the spin wave corresponding to $k = 0$ calculated from the dispersion relation in both the materials are almost the same, around 11 GHz, giving rise to propagating spin waves with minimum reflection. The wide range of frequencies excited corresponds to a range of wave vectors following the dispersion relationship. Following ref. [51], we extracted a dominating wavelength of 210 nm that corresponds to a wave vector (k) of 3x10$^7$ m$^{-1}$. Fig. 2(d) shows the spin wave transmission from the transmitter ME cell through the SWB to the detector ME cell and the frequency spectra along the length of the SWB, obtained from FFT of the *x*-component of the magnetization. Since the amplitude of the spin wave propagating through or detected at the detector ME cell depends on the voltage-induced net out-of-plane anisotropy $K$ of the detector ME cell (see eqn. 6 of supplementary information), for a fixed applied voltage at the detector, the presence of $\epsilon_{res}$ manifests itself as an increase in $K$ resulting in a higher damping of the propagating spin wave.

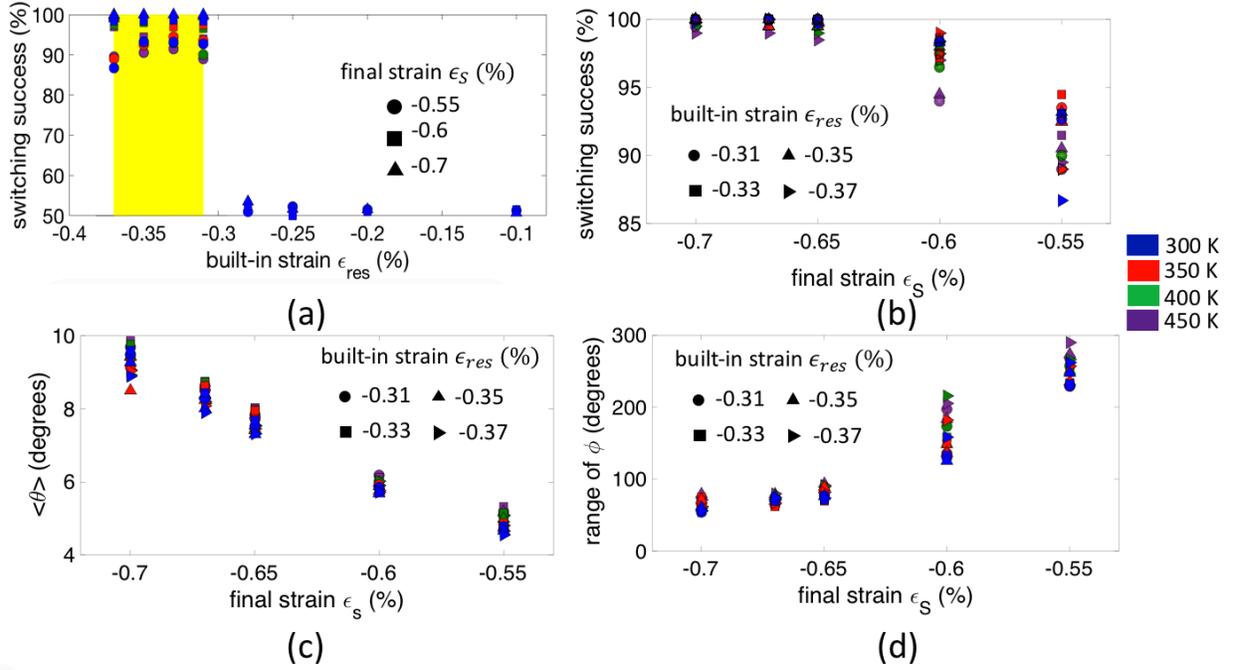

Figure 3: (a) Plot illustrating the dependence of the switching success on the built-in strain, (b) Impact of the final strain $\epsilon_s$ on the switching success which stems from the effect on the detected mean amplitude ($\theta$) and range of the phase ($\phi$) of the spin wave as shown in (c) and (d), respectively. Symbols illustrate different final strains $\epsilon_s$ in (a) and built-in strains in (b-d) while colors indicate different temperatures (300 K – 450 K).

Fig. 3(a) demonstrates the impact of $\epsilon_{res}$ on the switching success of the ME spin wave detector. We define the switching success as the probability of achieving an error-free logic functionality (buffer/inverter) in the presence of thermal noise. The narrow window of strain highlighted in the figure, within which the locations of the energy maxima and the saddle point interchanges, shows a dramatic increase in the switching success. It is intriguing to find that there is yet another parameter that affects the switching success within this range of built-in strain - the magnitude of the final strain $\epsilon_s = \epsilon_{res} + d_{31} V/t_{PZ}$ due to applied voltage for in-plane to out-of-plane magnetization switching. As shown in Fig. 3(b), the success rate increases with an increase in the magnitude of $\epsilon_s$ for all values of $\epsilon_{res}$. The impact of $\epsilon_s$ stems from two distinct effects. Firstly, the mean amplitude of the spin waves (<$\theta$>) excited by the transmitter ME cell increases with the magnitude of $\epsilon_s$ as depicted in Fig. 3(c). Secondly, the range of the detected phase of the spin wave (approximated as a $6\sigma$ deviation from the mean value) decreases with the increase in $|\epsilon_s|$ as shown in Fig. 3(d). In other words, the capability to have a correct detection of the phase of the spin wave increases with the magnitude of $\epsilon_s$ due to - (i) generation of higher amplitude spin waves resulting in a higher signal to noise ratio (SNR) at the point of detection, and (ii) decrease in the inherent thermal fluctuations of the detector ME cell, i.e., a decrease in the thermal noise floor. To investigate the robustness of our proposed scheme with respect to thermal noise, we include the effect of different temperatures ranging from 300 K to 450 K in Fig. 3. Overall, we see very little difference highlighting the robustness of the scheme relative to thermal noise.

**Exchange-spring system**

Next we explore another interesting and a more flexible option to tailor the energy landscape and the spin configuration by placing the PMA [Co(0.4)/Ni(0.8)]$_{10}$ SWB and the in-plane magnetized Co$_{60}$Fe$_{40}$ layer one on top of the other as illustrated in Figs. 4(a) and (b). Such a configuration, commonly referred to as exchange-spring [52,53], exhibits a much stronger exchange-coupling between the ME cell and PMA SWB compared to the earlier structure, and by taking advantage of the strong competition between the shape anisotropy of the Co$_{60}$Fe$_{40}$ layer (favoring in-plane magnetization) and the PMA of the [Co(0.4)/Ni(0.8)]$_{10}$ multilayer, a desired magnetization tilt angle can be achieved. Additionally, the strong interlayer exchange coupling forbids the out-of-plane +z direction to have the energy maxima as shown in Fig. 4(c), the condition we desire to achieve for thermal reliability. The impact of the change in the energy landscape has the same effect as explained earlier. We also find that the energy landscape and consequently the tilt angle can be varied by changing the thickness of Co$_{60}$Fe$_{40}$ layer ($t_{ME}$), and thus ensure the non-volatility of the magnetization states under zero applied voltage (see supplementary section S9).

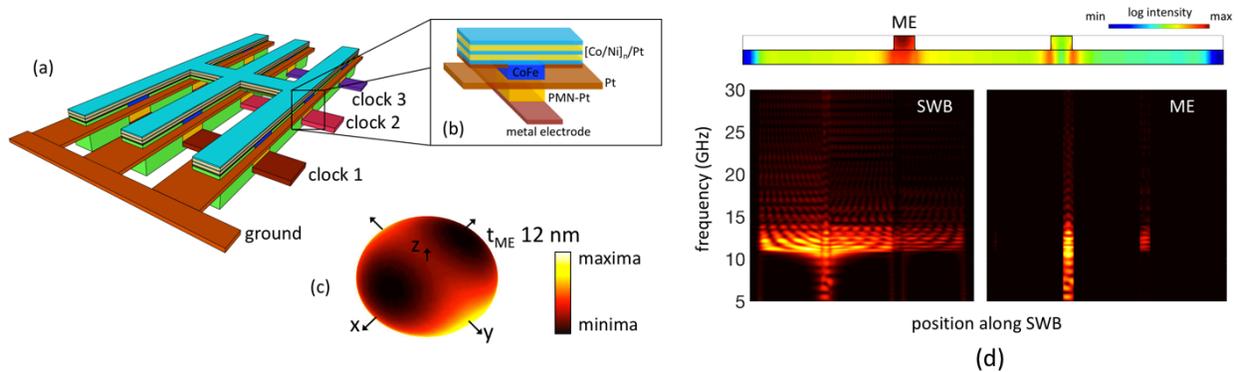

Figure 4: (a) Illustration of an alternative layout of a spin wave logic circuit, (b) Details of the main building blocks - ME cell and PMA SWB placed in a so-called exchange-spring configuration. (c) Energy landscape of the CoFe layer of the ME cell exchange coupled to the PMA Co/Ni SWB. (d) Spin wave transmission from the transmitter ME cell through the SWB to the detector ME cell and the frequency spectra along the length of the SWB, obtained from FFT of the *x*-component of the magnetization

The [Co/Ni] and the CoFe layers are coupled via volume exchange interaction as mentioned earlier. Similar to the case of built-in strain, voltage-driven strain-mediated magnetization switching of the transmitter ME cell excites spin waves with a wide range of frequencies. However, only those frequencies which are above the cutoff frequency of the [Co/Ni] multilayer and CoFe layer are allowed to penetrate and propagate through the SWB and ME cell. The values of $M_S$ and net out-of-plane $H_K$ for [Co/Ni] and exchange-coupled CoFe are very similar giving rise to almost same cut-off frequencies. Fig. 4(d) shows the spin wave transmission from the transmitter ME cell through the SWB to the detector ME cell and the frequency spectra along the length of the SWB showing the coupling of mode in [Co/Ni] to the CoFe layer. The strong inter-layer exchange coupling results in additional damping of the propagating spin waves resulting in a decrease in the amplitude of the propagating spin wave through the spin wave bus compared to the case without the ME cell.

Fig. 5(a) demonstrates the impact of $t_{ME}$ on the switching success of the ME spin wave detector. For relatively thin ME cell of around 8 nm, the reduction in the switching success can be attributed to the low energy barrier (less than $40k_BT$) between the "zero-voltage" canted magnetization states. In contrast to the narrow window of required built-in strain, here, the switching success increases with $t_{ME}$ and eventually saturates. This is because the condition for the energy maxima to be at +z direction is not enabled in all cases owing to the strong exchange coupling between the ME cell and the PMA SWB which prefers parallel spin alignment. We also see a dependence of the switching success on the applied voltage (Fig. 5(b)) which can be explained by looking at the dependence of the detected mean amplitude ($<\theta>$) and the range of the detected phase ($\phi$) of the spin wave on the applied voltage shown in Figs. 5(c) and (d), respectively. We also investigate the robustness of our proposed scheme with respect to thermal noise by including the effect of different temperatures ranging from 300 K to 450 K in Fig. 5.

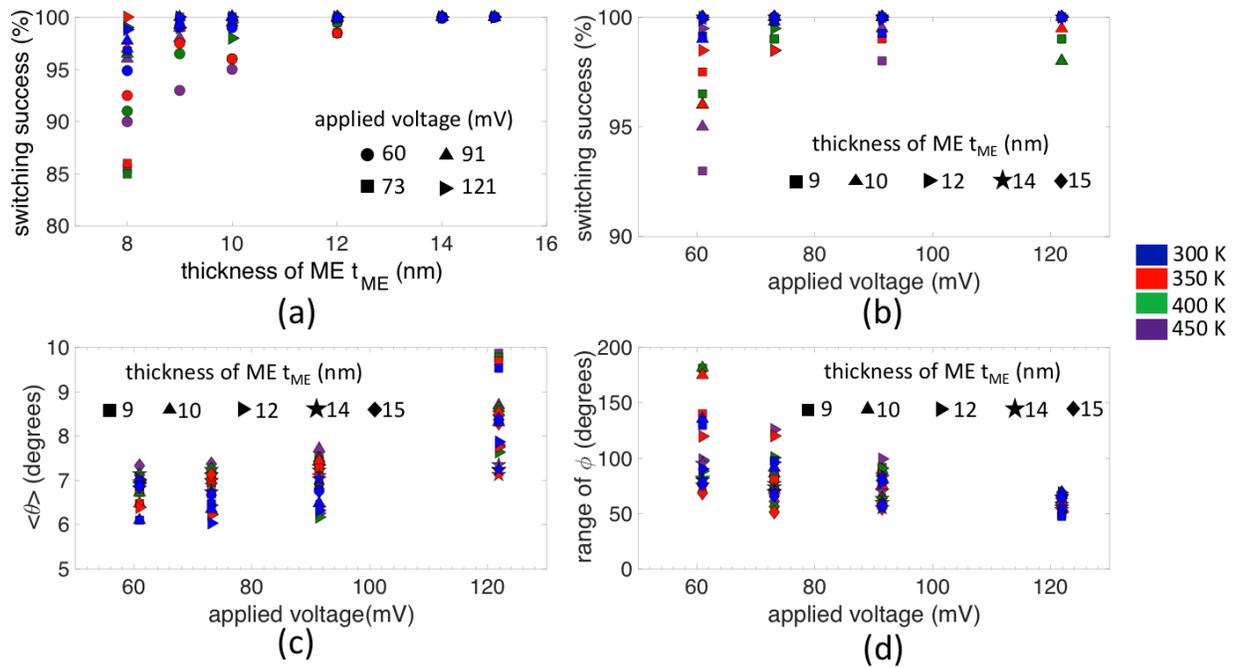

Figure 5: (a) Plot illustrating the dependence of the switching success on the thickness of the ME cell $t_{ME}$, (b) Impact of the applied voltage on the switching success which stems from the effect on the detected mean amplitude ($\theta$) and range of the phase ($\phi$) of the spin wave as shown in (c) and (d), respectively. Symbols illustrate different applied voltages in (a) and thickness of the ME cell $t_{ME}$ in (b-d) while colors indicate different temperatures (300 K – 450 K).

**Clocking error**

Establishing the fact that the proposed spin wave detection scheme is sensitive to the time of clocking, another error that enters into the picture and can have a significant impact on the reliability of the spin wave logic device is the clocking error. The clocking error can stem from sources such as clock jitter or transmitter-receiver clock skew. To study this error, we sweep the time of clocking normalized to the time period of the propagating spin waves and calculate the probability of error-free logic functionality at each clocking time. In addition to a change in the logic function of the device from an inverter to a buffer, we also observe a switching margin in the range of $T_{SW}/4$ to $T_{SW}/3$ within which an error-free logic functionality can be ensured. In our simulation for propagating spin waves with frequency around 11-13 GHz ($T_{SW} = 77 - 90\ ps$), we observe a switching margin of 20 – 30 ps. Assuming CMOS clocks operating in the frequency range of 3-5 GHz with 10% clocking error, we can expect to achieve such small clock margin although it may be challenging. Figs. 6(a) and (b) show the simulation results obtained for the two approaches mentioned earlier - built-in strain and exchange-spring, respectively.

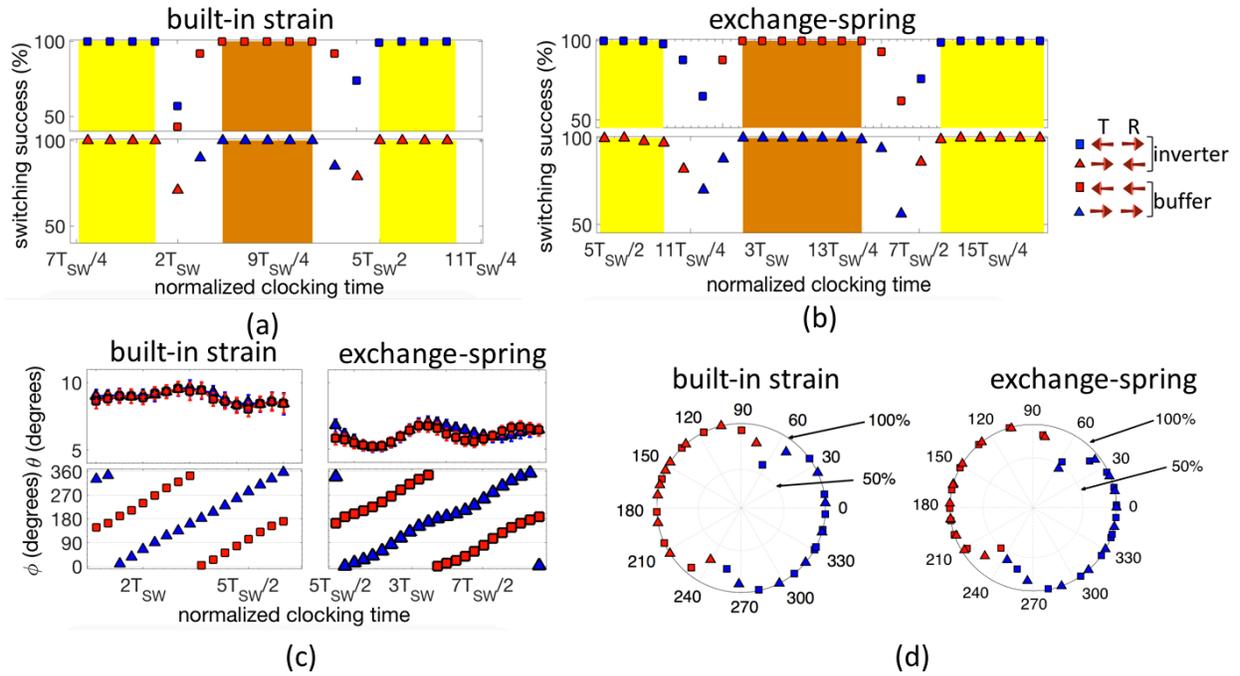

Figure 6: (a), (b) Plot illustrating the impact of the time of clocking on the switching success for both the case of built-in strain and exchange-spring, respectively. In addition to a change in the logic function of the device from an inverter to a buffer, we also observe a switching margin in the range of $T_{SW}/4$ to $T_{SW}/3$ within which an error-free logic functionality can be ensured. (c) Detected amplitude ($\theta$) and phase ($\phi$) of the spin wave as function of the time of clocking. Note the error bars indicate the deviation ($\sigma_\phi$) from the mean value due to the presence of thermal noise. (d) Switching success as a function of the detected mean phase. An error-free logic functionality is achieved if the detected phase falls within the window from 280° through 0 to 20°, i.e. 100°, or from 100° to 200°.

To better understand the results, it is essential to look into the mean values of the detected amplitude ($<\theta>$) and phase ($<\phi>$) of the spin wave. A change in the time of clocking results in a change in the detected phase which dictates the direction of magnetization relaxation, in other words the functionality of the device (Fig. 6(c)). Next, we plot the switching success as a function of the detected mean phase ($<\phi>$) as shown in Fig. 6(d) for both the approaches. It is seen that if $<\phi>$ lies with the window from 280° through 0 to 20°, i.e. a 100° margin, we end up with an error free switching of magnetization to the +x direction while the window from 100° to 200°, also 100° margin, results in an error free switching to -x. The reason for the tilt in the distribution (asymmetric with respect to the line joining 90° and 270°) stems from the energy landscape and the constant energy trajectories (see supplementary section S10 for details). We would like to emphasize that our results are in contrast to prior work that assumed the binary output (logic 1 or 0) would depend on the phase of the incoming spin wave falling in the range of -90° to 90 or 90° to 270° respectively [54]. To have thermally reliable deterministic switching, the detected phase should fall within the window from 280° through 0 to 20° or from 100° to 200°.

## Discussion

Based on what has been described until now, it is possible to set forth a design rule for ensuring the thermal reliability of the spin wave logic device. As highlighted in Fig. 7, combining results from both the approaches, a high switching success and error-free logic functionality can be ensured if the amplitude of the detected spin wave ($<\theta>$) remains higher than a threshold value of around 6° and the detected phase falls within the window from 280° through 0 to 20° or from 100° to 200° with a maximum allowable $\phi$ range of around 100°. Note that the increase in the magnetic damping of the spin wave channel from the simulation value of 0.01 used here due to extrinsic contributions like sample inhomogeneity will result in a decrease of the mean amplitude of the spin wave ($<\theta>$) at the detector ME cell. The lowering of $<\theta>$ below the critical threshold for the case of enhanced magnetic damping ~ 0.1 will result in a decrease of the switching success of the ME spin wave detector. However, we can still ensure the thermal reliability and error-free logic functionality by shortening the length of the spin wave channel which results in a higher spin wave amplitude ($<\theta>$) at the detector ME cell.

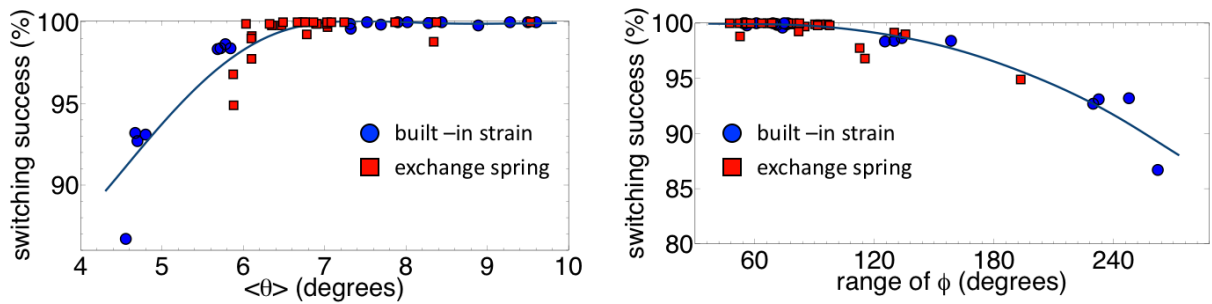

Figure 7: Dependence of the switching success on the detected amplitude ($\theta$) and phase ($\phi$) of the spin wave, combining results from both the approached. A high switching success and error-free logic functionality can be ensured if the amplitude of the detected spin wave ($<\theta>$) remains higher than a threshold value of around 6° and the detected phase falls within the window of 280° through 0 to 20° or 100° to 200° with a maximum allowable $\phi$ range of around 100°.

In conclusion, we have realistically assessed the possibility of developing magnonic logic device that meets all the requirements for logic circuits and is robust to thermal noise and variability.

We started by identifying suitable materials for the ME cell and PMA SWB using experimentally demonstrated parameters. The chosen materials for the SWB and ME cell are appealing owing to their ability to sustain a propagating spin wave via low damping and reduced PMA, high product of coupling coefficient to enable ultra-low power dissipation, ease of fabrication and material compatibility. Next, we explored the impact of thermal noise on the magnetization dynamics of the ME cell and in terms of the phase noise of the spin waves. A salient feature of this work is the translation of the "theoretical idea" of phase-dependent switching of the spin wave detector [46] to a "practical realization" of a thermally reliable magnonic device. We identified two viable options: built-in strain and exchange spring system, both relying on a change in the energy landscape and a phase-detection scheme utilizing saddle-point based magnetization switching.

Furthermore, the work revolves around a unique scheme of building memory cells on top of the logic gate to retain input data, trigger spin waves, and read out the result. Such a proposal enables the realization of two primary logic gates, inverters and majority gates, that lie at the heart of wave-based computing and together with the new emerging novel logic synthesis technique [55] can open up and enable the true potential of the field of spin waves. Overall, this work addresses a very critical question: "Can spin wave devices work in reality?" and provide a solid platform towards the practical realization of an error-free ultra-low power spin wave logic device. We also believe that our simulations can be a guide for the development of an error-free magnonic logic device and will inspire more future experimental work in this field.

## Methods
### Micromagnetic simulation

We performed micromagnetic simulation using the Object Oriented Micromagnetic Framework (OOMMF)[56] that numerically solves the stochastic Landau-Lifshitz-Gilbert equation augmented with thermal noise. We chose a 100 nm long [Co(0.4)/Ni(0.8)]$_{10}$ multilayer as the SWB having a total thickness of 12 nm. The width was fixed at 40 nm. The effective PMA, saturation magnetization and exchange stiffness for the multilayer were calculated as K = 0.4 MJ/m$^3$, M$_s$ = 790 kA/m and A = 16 pJ/m, respectively (see supplementary section S2 for detailed calculation). Experimentally determined values of the gilbert damping constant $\alpha$ in [Co/Ni] multilayers have varied from 0.015 to around 0.1. Recent works by Haertinger et. al.[39], Mizukami et. al. [40], Shaw et. al. [41] and Beaujour et. al. [42] have reported a rather low $\alpha$ depending on the layer thickness and bilayer periodicity. From the plots of $\alpha$ as a function of the thickness of the Co layer and bilayer period presented in[40], we estimated a value of around 0.01 for our [Co/Ni] multilayer spin wave bus. For the ME cell we used the material parameters corresponding to Co$_{0.6}$Fe$_{0.4}$: saturation magnetization M$_s$ = 800 kA/m [57], exchange stiffness A = 20 pJ/m, gilbert damping $\alpha$ = 0.027 [57], magnetostrictive coefficient $\lambda$ = 200 ppm[45], Young's modulus Y = 200 GPa. The lateral dimension of the Co$_{0.6}$Fe$_{0.4}$ layer was taken as 80 nm x 40 nm. For the case of built-in strain where we used a ME cell embedded within the spin wave bus, the thickness was same as that of the SWB, ie, 12 nm, while for the case of exchange-spring, the thickness was varied between 8 and 15 nm. The interface between [Co/Ni] and CoFe is modelled using simple volume exchange energy where the exchange stiffness at the interface is taken as the average value of that for [Co/Ni] and CoFe $A_{interface} = \frac{A_{Co/Ni} + A_{CoFe}}{2}$.

The full 3D stochastic micromagnetic simulation can be computationally demanding when performing Monte Carlo simulations for thermal reliability. Hence, we resort to a 1D micromagnetic simulation for the case of built-in strain (discretization only along the length with cell size $\delta_x$ = 5 nm) and a 2D simulation for exchange-spring system (discretization along the length and thickness with cell size $\delta_x$ = 5 nm and $\delta_z$ = 2-3 nm). Absorbing boundary condition with higher damping was employed at the ends of the SWB to avoid reflections. For thermal reliability, we performed 1000 Monte Carlo simulations for each data point to determine the probability of error-free logic functionality. A comparison with full 3D simulation is provided in supplementary section S11 for a limited design space which shows good qualitative agreement between the two in terms of switching success as a function of the clocking time and detected phase. Since an error-rate calculation using brute force micromagnetic simulation can be computationally exhaustive, we have restricted simulations to only 1000 trials that gave an error rate of < $10^{-3}$. The recently developed "rare-event enhancement" (REE) technique for micromagnetics [58] cannot be trivially applied to our fast picosecond magnetization switching dynamics. Hence, to capture the extreme tails of error-rate, we use a less computationally intensive equivalent single domain stochastic LLG simulation for the SW detector[46] and the "rare-event enhancement" (REE) technique to reach an error-rate of less than $10^{-9}$. The details are provided in supplementary section S12.

## Acknowledgement

The authors would like to thank Dr. R. H. Victora at University of Minnesota and Dr. J. Shaw of NIST for useful discussion and insight on damping phenomena.

## Author Contributions

S.D., D.E.N, S.M., I.A.Y. and A.N. developed the main idea. S.D. performed the simulations. All authors discussed the results, agreed to the conclusions of the paper and contributed to the writing of the manuscript.

**Competing financial interests:** The authors declare no competing financial interests

# Overcoming thermal noise in non-volatile spin wave logic

# – Supplementary information

Sourav Dutta[*,1], Dmitri E. Nikonov[2], Sasikanth Manipatruni[2], Ian A. Young[2], and Azad Naeemi[1]

[1]School of Electrical and Computer Engineering, Georgia Institute of Technology, Atlanta, GA 30332 USA

[2]Components Research, Intel Corporation, Hillsboro, OR 97124 USA

**Contents**
S1. Comparison with alternative spin configuration of ME-SWB system
S2. Calculation of perpendicular magnetic anisotropy (PMA) of Spin Wave Bus
S3. Comparison with other PMA SWB
S4. Comparison of material parameters for ME cell
S5. Ultrahigh strain and strain relaxation
S6. Mathematic expression for ME effect
S7. Device working principle
S8.  Comparison with voltage-controlled magnetic anisotropy (VCMA) effect
S9. Non-volatility and magnetization tilting in the presence of built-in strain and in exchange-spring structure
S10. Asymmetric tilted distribution of switching success as a function of detected $< \phi >$
S11. Comparison between 1D and full 3D micromagnetic simulation
S12. Approximate estimation of error rate

**S1. Comparison with alternative spin configuration of ME-SWB system**

**S1.1. PMA SWB – in-plane ME cell**
In this paper, we explore the system of PMA spin wave bus (SWB) and ME cell with stable magnetization states along the long axis (+/- x) as shown in Fig. 1(a). Note that the choice of mutually orthogonal spin configuration of SWB and ME cell stems from the requirements of non-volatility and non-reciprocity [1]. Applying a voltage aligns the magnetization of the ME cell with that of the SWB allowing the spin waves to arrive and subsequently get detected. This working principle can be extended to in-plane magnetized SWB.

**S1.2. In-plane SWB – in-plane ME cell**
Fig. 1(b) shows a magnetostatic surface SWB magnetized along the in-plane hard axis (y) and the corresponding mutually orthogonal magnetization of the ME cell. The magnetization of the ME cell would be energetically favored along the x-axis due to the inherent shape-anisotropy and the presence of a possible magnetocrystalline anisotropy. Application of a voltage in this scenario across a (011) cut ferroelectric layer can create an anisotropic strain[2] to rotate the easy axis from x to y and switch the magnetization. Note that since the ME cell would have a naturally occurring saddle point along the y axis in the absence of any built-in strain, the saddle point based phase detection of the spin waves would be valid in the scenario. However, a major drawback is the requirement of an external biasing field to create the transverse magnetization in SWB nanowire.

### S1.3. In-plane SWB – PMA ME cell

Fig. 1(c) shows a longitudinally magnetized backward volume SWB along with a PMA ME cell. The PMA ME cell can be obtained by using the surface anisotropy at the interface of a magnetostrictive layer and a metal/ oxide (Ni/Cu [3], CoFe/MgO [4] etc.) or large in-plane built-in strain like Ni/BaTiO$_3$ [5]. An applied voltage can create an in-plane isotropic or anisotropic strain to lower the PMA and cause in-plane switching of magnetization towards the x-axis (favored due to shape anisotropy). Note that for a PMA magnet, the saddle point is located at the x-axis and hence the saddle point based phase detection of the spin waves would also be valid in the scenario. However, as mentioned earlier in the main text, the broken translational symmetry and anisotropic dispersion relation of the backward volume spin waves can give rise to scattering processes where the waves interfere.

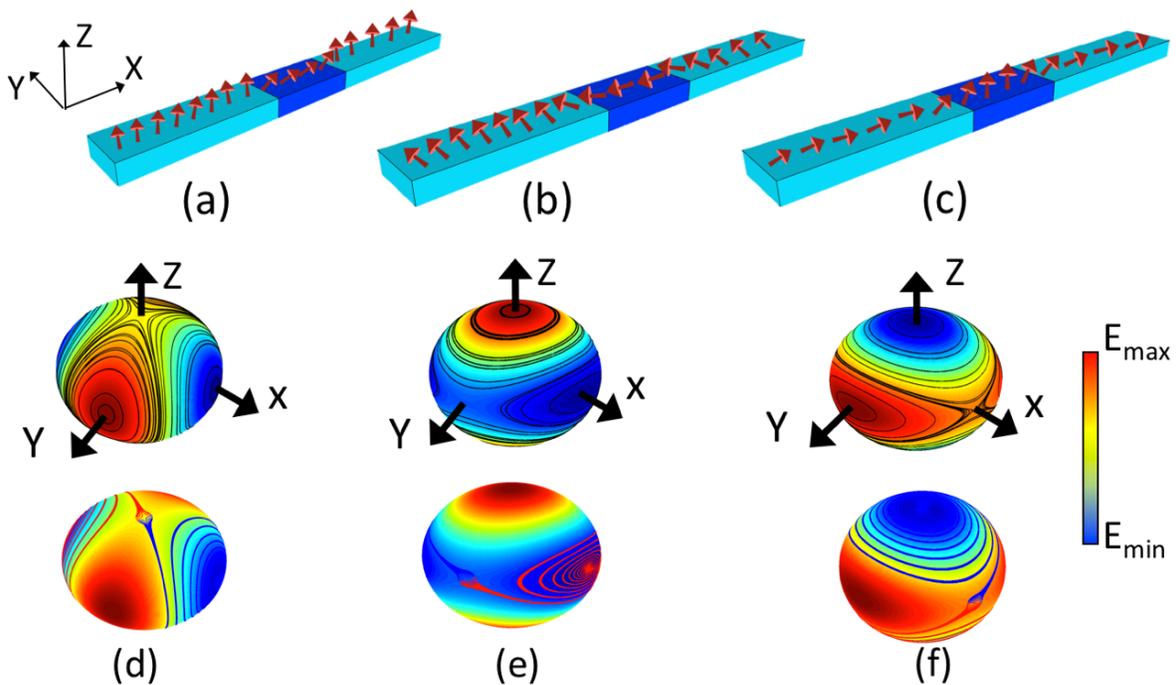

Figure 1. Illustration of the possible spin configuration of the ME-SWB system.

## S2. Calculation of perpendicular magnetic anisotropy (PMA) of Spin Wave Bus

The origin of magnetic anisotropy can be attributed to two mechanisms: (a) the long range magnetic dipolar interaction which gives rise to the shape anisotropy, and (b) the spin-orbit interaction which gives rise to magnetocrystalline anisotropy and magnetoelastic anisotropy. The perpendicular magnetic anisotropy (PMA) arises from this spin-orbit interaction at the interface which has a lowered symmetry, and hence behaves differently from bulk magnetic anisotropy. Specifically, theoretical prediction by Daalderop et. al. [6] in Co/Ni multilayer system has revealed the PMA to be arising from the spin-orbit interaction of states with $d_{x^2-y^2}$ and $d_{xy}$ character present close to the Fermi level.

Phenomenologically, the total effective magnetic anisotropy can be separated into a volume contribution $K_V$ and a surface contribution $K_S$ and can be expressed as a function of the Co and Ni layer thicknesses $t_{Co}$ and $t_{Ni}$ and number of bilayer repetitions n as[7-10]:

$$K^{eff}D = K_V^{Co}t_{Co} + K_V^{Ni}t_{Ni} + 2K_S^{Co/Ni} + \frac{1}{n}\left[K_S^{Co/Pt} + K_S^{Ni/Pt} - K_S^{Co/Ni}\right] \quad (1)$$

where D is the bilayer thickness (D = $t_{Co}$ + $t_{Ni}$ = $(1 + \alpha)t_{Co}$), $\alpha$ = $t_{Ni}/t_{Co}$ is the thickness ratio and $K_V^{Co}$ and $K_V^{Ni}$ are the volume anisotropies of Co and Ni layers, respectively. $K_S^{Co/Ni}$, $K_S^{Co/Pt}$ and $K_S^{Ni/Pt}$ are the interface anisotropies of Co/Ni, bottom Co/Pt and top Ni/Cap interfaces, respectively, considering the deposition of the Co/Ni multilayer on an underlayer of Pt and capped with a top capping layer, say Ta. Neglecting the effect of top cap layer, and assuming the following parameters: $K_V^{Co}$ = -1 MJ/m$^3$, $K_V^{Ni}$ = -0.12 MJ/m$^3$, $K_S^{Co/Ni}$ = 0.22 mJ/m$^2$, $K_S^{Co/Pt}$ = 0.88 mJ/m$^2$, we calculate the PMA of the multilayer as a function of the Co layer thickness $t_{Co}$ for a fixed number of bilayers n = 10 as shown in Fig. 1(a,b) and as a function of n for a fixed thickness ratio $\alpha$ of 2.

The effective saturation magnetization of the multilayer is calculated as
$$M_S D = M_S^{Co}t_{Co} + M_S^{Ni}t_{Ni} \quad (2)$$
where $M_S^{Co}$ and $M_S^{Ni}$ are the saturation magnetization of the Co and Ni layers with assumed valued of 1.4 MA/m and 485 kA/m, respectively.

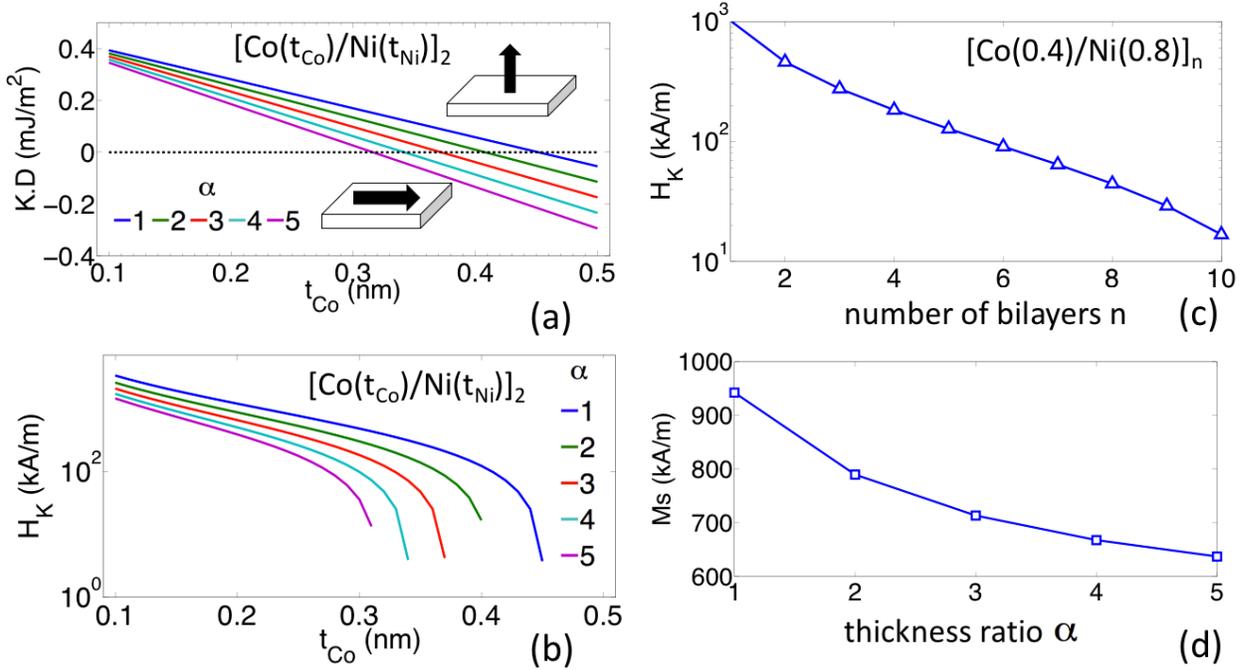

Figure 2: (a) Variation of K.D as function of the thickness of the Co layer $t_{Co}$ in the Co/Ni multilayer stack. A positive value indicates a PMA case while negative indicates in-plane magnetization. (b), (c) Variation of the anisotropy field HK with the thickness of the Co layer $t_{Co}$ and the number of bilayers n. (d) Variation of saturation magnetization of multilayer stack with the thickness ratio $\alpha$.

## S3. Comparison with other PMA SWB

The usage of surface magnetic anisotropy has been proposed to provide an out-of-plane biasing in single layer ultrathin SWB (~ 1nm) [11,12]. However, we anticipate such thin spin wave channel to be highly prone to channel noise and give rise to phase noise of propagating spin waves.

## S4. Comparison of material parameters for ME cell

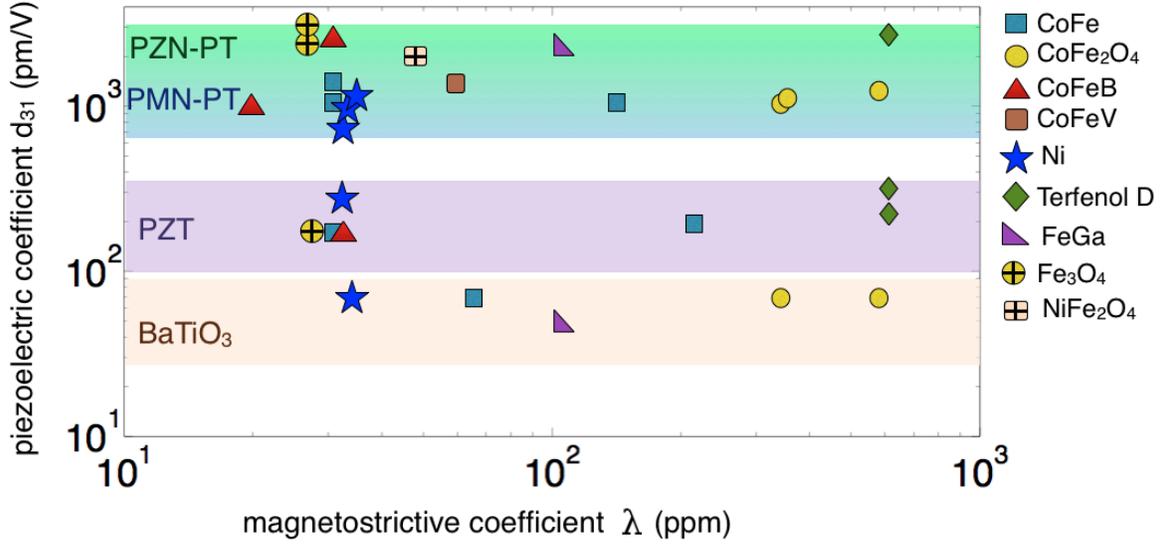

Figure 3: Map of the magnetostrictive coefficient $\lambda$ of the magnetic layer and piezoelectric coefficient $d_{31}$ of the piezoelectric/ferroelectric layer along with their compatibility for some of the demonstrated stacks.

For minimizing power dissipation, the target piezoelectric material must possess a high piezoelectric coefficient ($d_{31}$) while the magnetic layer must display a high magnetostrictive coefficient ($\lambda$) simultaneously. Fig. 3 shows a comprehensive map of the piezoelectric and magnetostrictive coefficients for a wide range of material including their compatibility. Nickel (Ni) [2,5,13-16] has been extensively used experimentally to demonstrate magnetoelectric effect, however it displays a low magnetostrictive coefficient ($\lambda \sim -32$ ppm). $CoFe_2O_4$ [17-23] offers an order of magnitude improved $\lambda$, however it may provide a lower thermal stability/reliability owing to low saturation magnetization. Similar high magnetostriction of $\lambda = 150$ ppm has been reported in equiatomic composition of $Co_{0.5}Fe_{0.5}$ [24,25]. Recent work by Hunter et. al. [26] has reported an enhancement of magnetostriction at the (fcc+bcc)/bcc phase boundary with effective $\lambda$ as high as 260 ppm. Alternative materials include low magnetostrictive CoFeB [27-29], CoFeV [30], $Fe_3O_4$ [31], $NiFe_2O_4$ [32] and high magnetostrictive $Fe_{0.8}Ga_{0.2}$ [33-36] ($\lambda > 250$ ppm) and highest magnetostriciton observed in $Tb_xDy_{1-x}Fe_2$ [37-39]. In this work, we consider $Co_{0.6}Fe_{0.4}$ with $\lambda = 200$ ppm. The comparison in Fig. 3 shows that, with the chosen combination of piezoelectric PMN-PT and the ferromagnet CoFe, one can reach a high product of coupling coefficient. An added advantage of our choice is a much more mature fabrication process for CoFe compared with that of Terfenol-D.

## S5. Ultrahigh strain and strain relaxation

In contrast to polycrystalline materials like Pb(Zr,Ti)$O_3$ (PZT), relaxor based ferroelectric single crystals like (Pb(Zn$_{1/3}$Nb$_{2/3}$)$O_3$)-[PbTi$O_3$] (PZN-PT) and (Pb(Mg$_{1/3}$Nb$_{2/3}$)$O_3$)-[PbTi$O_3$](PMN-PT) do not require morphotropic phase boundary conditions for exhibiting ultrahigh piezoelectric strain [40]. <001> oriented relaxor based rhombohedral crystals such as (1-x)PZN-PT (x < 9 %) and (1-x)PMN-PT (x < 35 %) are known to demonstrate ultrahigh piezoelectric coefficient $d_{33}$ and strains of 0.6 % - 0.8 % with applied electric field less than the dielectric breakdown limit [40]. The reason for such high strain is presumed to be associated with an electric field induced rhombohedral-tetragonal phase transition.

Substrate induced clamping can drastically reduce the piezoelectric response of thin ferroelectric films from their bulk value [41]. As such we assume the thickness of the PMN-PT film to be at least greater than 30-50 nm. We also assume the lateral size of the PMN-PT film to be larger than the thickness to allow the formation of an in-plane isotropic biaxial strain instead of anisotropic strain [42]. It has been demonstrated that a high strain relaxation of up to 90% can occur in thick ferroelectric films (thickness > lateral dimensions) [42]. As such we limit the thickness of the CoFe layer of the ME cell to around 12 nm.

## S6. Mathematic expression for ME effect

The magnetoelastic energy [43] describing the coupling between the magnetization and the strains can be written in first-order approximation as:

$$E_{ME} = -\frac{3}{2}\lambda Y \left[\left(m_x^2 - \frac{1}{3}\right)\epsilon_{xx} + \left(m_y^2 - \frac{1}{3}\right)\epsilon_{yy} + \left(m_z^2 - \frac{1}{3}\right)\epsilon_{zz}\right] \quad (3)$$

where $m_i$ (i=x, y, z) are the direction cosines of magnetization $\vec{M}$, $\lambda$ is the magnetostrictive constant, Y is the Young's modulus and $\epsilon_{xx}$, $\epsilon_{yy}$ and $\epsilon_{zz}$ are the strains in the x, y and z directions.

The application of an out-of-plane electric field $E_Z$ across the (001) cut ferroelectric PMN-PT layer creates an isotropic bi-axial in-plane strain [23,44] given by

$$\epsilon_{xx} = \epsilon_{yy} = \epsilon_s = \epsilon_{res} + d_{31}E_Z \quad (4)$$

where $\epsilon_{res}$ represents the in-plane built-in strain and $d_{31}$ represents the piezoelectric coefficients.

The energy expression can be further reduced to [1,5]

$$E_{ME} = -\frac{3}{2}\lambda Y \left[\left(m_x^2 - \frac{1}{3}\right)\epsilon_{xx} + \left(m_y^2 - \frac{1}{3}\right)\epsilon_{yy}\right] = \frac{3}{2}\lambda Y \left(m_Z^2 - \frac{1}{3}\right)\epsilon_S \quad (5)$$

The equivalent out-of-plane strain-induced anisotropy can be calculated as

$$K = -\frac{3}{2}\lambda Y \epsilon_S = -\frac{3}{2}\lambda Y(\epsilon_{res} + d_{31}E_Z) = -\frac{3}{2}\lambda Y\left(\epsilon_{res} + d_{31}\frac{V}{t_{PZ}}\right) \qquad (6)$$

where V and $t_{PZ}$ are the voltage applied across and the thickness of the piezoelectric layer respectively. For a stable out-of-plane magnetic configuration, K should be sufficient to overcome the out-of-plane shape anisotropy. The required voltage to create this strain-induced anisotropy is given by

$$V = -\left(\frac{2K}{3\lambda Y} + \epsilon_{res}\right)\frac{t_{PZ}}{d_{31}} \;. \qquad (7)$$

## S7. Device working principle

Under zero-applied voltage, the magnetization of the ME cell stays in-plane storing either a logic 1 (+x magnetization, Fig. 4(a)) or logic 0 (-x magnetization, Fig. 4(b)). Applying an out-of-plane electric field across the thickness of a (001) oriented ferroelectric or piezoelectric layer (poled in the perpendicular direction and having in-plane isotropic properties) causes an in-plane biaxial strain that gets coupled to the overlaying ferromagnetic layer through the interface and thin Pt electrode. An up to 90° magnetic easy axis rotation can be achieved as the isotropic in-plane strain surpasses a critical limit causing a voltage-induced strain-mediated out-of-plane anisotropy and subsequently magnetization switching. Such in-plane to out-of-plane magnetization switching dynamics can be used to excite spin waves, with the information encoded into the phase of the waves [1]. A +x to +z magnetization switching creates a spin wave with zero phase (Fig. 4(a)), while a -x to +z switching creates spin waves with opposite π phase (Fig. 4(b)). While the transmitter ME cell switches to excite spin waves, the detector ME cell is held in the out-of-plane meta-stable state via application of voltage until the incoming spin waves arrive. Upon arrival, the voltage is switched off causing a phase-dependent out-of-plane to in-plane magnetization switching. Interestingly, depending upon the time the voltage is switched off (time of clocking), we end up with the detector ME cell's magnetization falling either in the +x or -x direction. In other words, we can define the logic function of the SW device (buffer or inverter) simply by choosing the appropriate time of clocking.

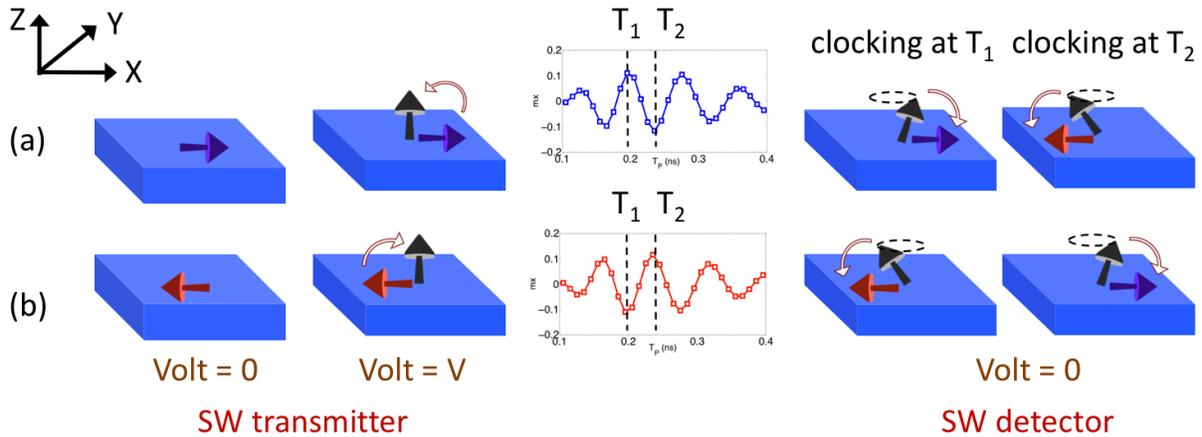

Figure 4: Working principle of the spin wave logic device.

## S8. Comparison with voltage-controlled magnetic anisotropy (VCMA) effect

An applied voltage across the interface of an oxide and ferromagnet, typically MgO and CoFeB, can vary the charge density at the interface giving rise to an alteration in the surface magnetic anisotropy. Similar to magnetostriction, VCMA can give rise to a change in the magnetic easy axis with upto 90° switching of magnetization. However, with the currently known materials (Fe/MgO [45], CoFe/MgO [4] and CoFeB/MgO [46-48]), surface anisotropy displays orders of magnitude lower magnetoelectric coefficient compared to magnetostriction [49]. Hence, while the device operation can still be performed with VMCA, the energy dissipation would be considerably higher.

## S9. Non-volatility and magnetization tilting in the presence of built-in strain and in exchange-spring structure

The presence of a built-in strain (less than the critical value) gives rise to a small perpendicular anisotropy $K = \frac{3}{2}\lambda Y \epsilon_{res}$ less than the shape anisotropy. The competition between the shape anisotropy favoring in-plane magnetization and PMA favoring out-of-plane configuration gives rise to a tilting of the magnetization from its stable in-plane configuration under zero-applied voltage as shown in Fig. 5(a). Beyond the critical strain of -0.48%, the magnetization goes out-of-plane thus losing the non-volatility of the zero-voltage magnetization states. The drop in the energy barrier as a function of the applied in-plane strain is also shown.

A more gradual change in the magnetization tilt angle as a function of the thickness of the CoFe layer is seen in the case of the exchange-spring system owing to the strong interlayer exchange coupling between the in-plane magnetized CoFe and PMA [Co/Ni] multilayer (Fig. 5(b)). Note that the energy barrier between the zero- voltage magnetization states drop to below 40k$_B$T for thickness less than 9 nm which explains the drop in the switching success of thinner ME cell magnet highlighted in the main text.

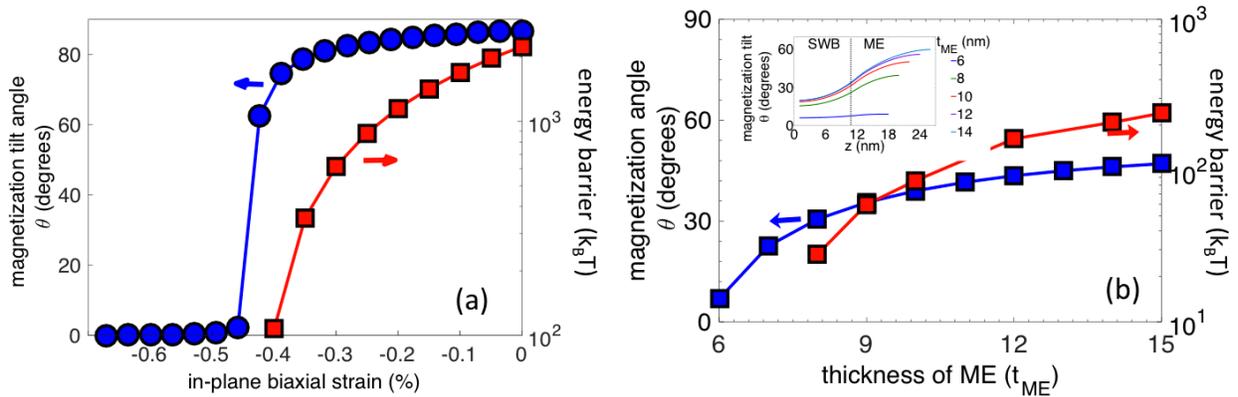

Figure 5: Magnetization tilt angle from the out-of-plane axis and the energy barrier between the stable zero-voltage magnetic states as a function of the (a) in-plane biaxial strain, and (b) thickness of the CoFe layer of the ME cell exchange coupled to the PMA SWB (Inset shows the magnetization profile or the tilt angle through the thickness of the PMA SWB-ME cell exchange-spring structure).

## S10. Asymmetric tilted distribution of switching success as a function of detected $<\phi>$

As shown in Fig. 6(d) of the main text, the distribution of the switching success as a function of the detected phase of the spin wave ($<\phi>$) is asymmetric with respect to the y-axis (line joining 90° and 270°). This tilted distribution can be explained by examining the energy landscape and the corresponding constant energy trajectories in the presence of a small built-in strain or exchange-spring system as shown in Fig. 6(a). The constant energy orbits go clockwise around the energy maxima and anti-clockwise around the energy minima [50]. In the presence of damping, the dynamical evolution of magnetization closely follows the orbits, resulting in magnetization switching trajectories as shown in Fig. 6(b). Note that the distribution of the initial angles of the magnet, which will dictate switching to either +x or -x direction, is now not centered around the 0° or 180° (x axis) but around the separatrix which separates the two types of constant energy trajectories (high energy orbits around maxima and low energy orbits around minima). The resultant tilted distribution under zero-thermal noise is shown in Fig. 6(c). Upon adding the thermal noise, the angles in the highlighted zone at the boundary gives rise to non-deterministic switching probability resulting in a lower switching success as shown Fig. 6(d) of the main text.

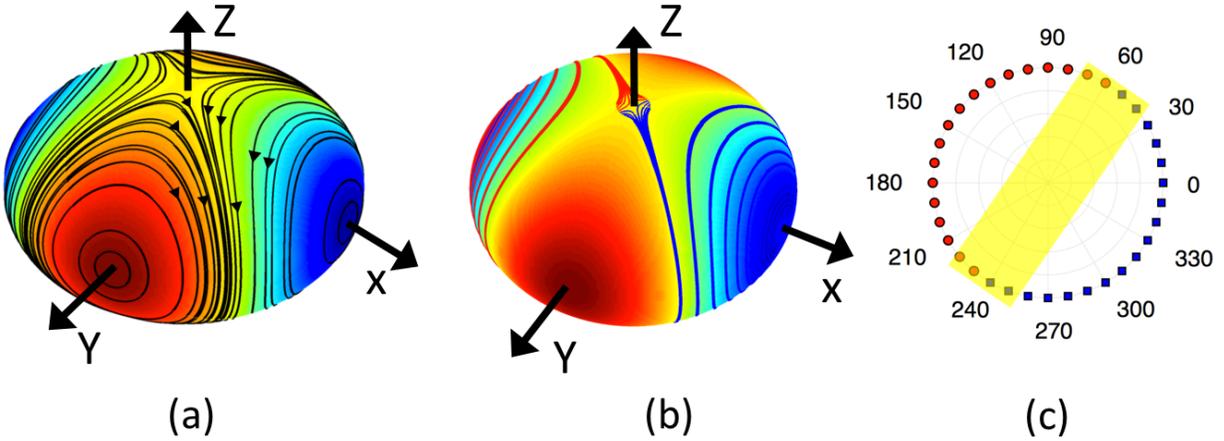

Figure 6: (a) Energy landscape and constant energy trajectories of the nanomagnet in the presence of a small built-in strain or exchange-spring system

## S11. Comparison between 1D and full 3D micromagnetic simulation

The full 3D stochastic micromagnetic simulation can be computationally demanding when performing Monte Carlo simulations for thermal reliability. Hence, we resort to a 1D micromagnetic simulation for the case of built-in strain (discretization only along the length) and a 2D simulation for exchange-spring system (discretization along the length and thickness). Such an approximation holds when considering a relatively narrow SWB where the spin waves are uniformly excited along the width of the SWB and propagate only along the length and the higher order width modes are not excited. To further corroborate our assumption, we compare our 1D simulation for built-in strain against the full 3D simulation for a clocking time sweep between 205 ps and 250 ps and show good qualitative agreement between the two in terms of switching success as a function of the clocking time (fig. 7a) and detected phase (fig. 7b). Note that, since we define the magnetization of the ME cell as a spatially averaged quantity, the exactly values of the detected amplitude and phase as a function of time varies a little between the two approaches hence the shift in Fig. 7(a). The result for error-free logic functionality achieved if the detected phase falls within the window from 280° through 0 to 20°, i.e. 100°, or from 100° to 200°. still holds when performing 3D simulation (Fig. 7(b)).

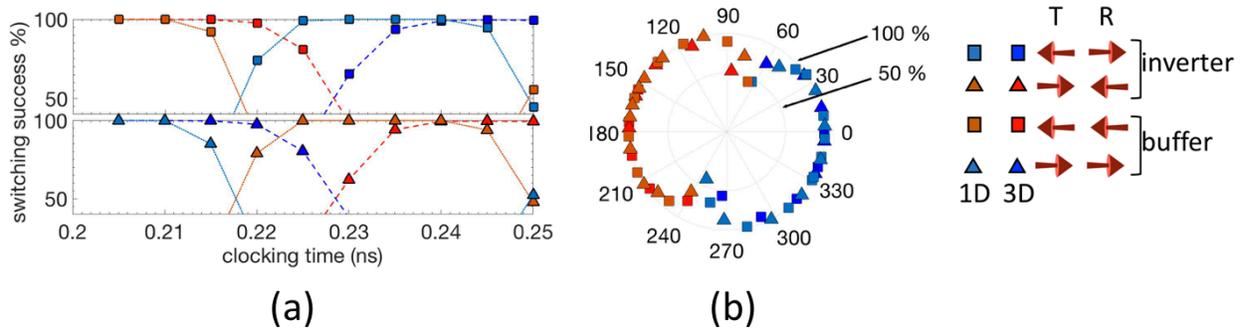

Figure 7. Comparison between 1D and full 3D micromagnetic simulations for switching success as a function of the clocking time (a) and detected phase (b)

## S12. Approximate estimation of error rate

Estimation of error rate is essential for ensuring the thermal reliability of a device. As mentioned in the methods of the main text, we 1000 Monte Carlo micromagnetic simulations in OOMMF for each data point to determine the probability of error-free logic functionality. Using such brute force technique to compute error rate below $10^{-3}$ can become computationally demanding and extreme tails of error-rate cannot be captured in this way. Also, the recently developed "rare-event enhancement" (REE) technique for micromagnetics [51] cannot be trivially applied to our fast picosecond magnetization switching dynamics. Hence, we resort to an equivalent single domain approach of modeling the SW detector [52]. Assuming a distribution of the initial angles $\theta$ and $\phi$ as shown in the insets of Fig. 8 to mimic the effect of arriving spin wave, we let the single domain magnet fall towards an energy minimum. Performing single domain stochastic LLG simulation gave us an error-rate of less than $10^{-5}$ for up to $10^5$ brute force trials. We further attempted to capture the extreme tails of error-rate by using the REE technique that artificially enhances the rate of occurrence of low-probability events while proportionately reducing their weights to reach an error-rate of less than $10^{-9}$ as shown in Fig. 8. An alternative approach is to use the Fokker-Planck equation. However, note that still a single domain approximation has to be made while using the Fokker-Planck method. Also note that unlike the case of a PMA magnet, the approximate analytical expression or numerical solution of a 1-D Fokker-Planck cannot be used and one has to resort to a numerical solution of a 2-D Fokker-Planck using FDM or FEM method and is beyond the scope of this work.

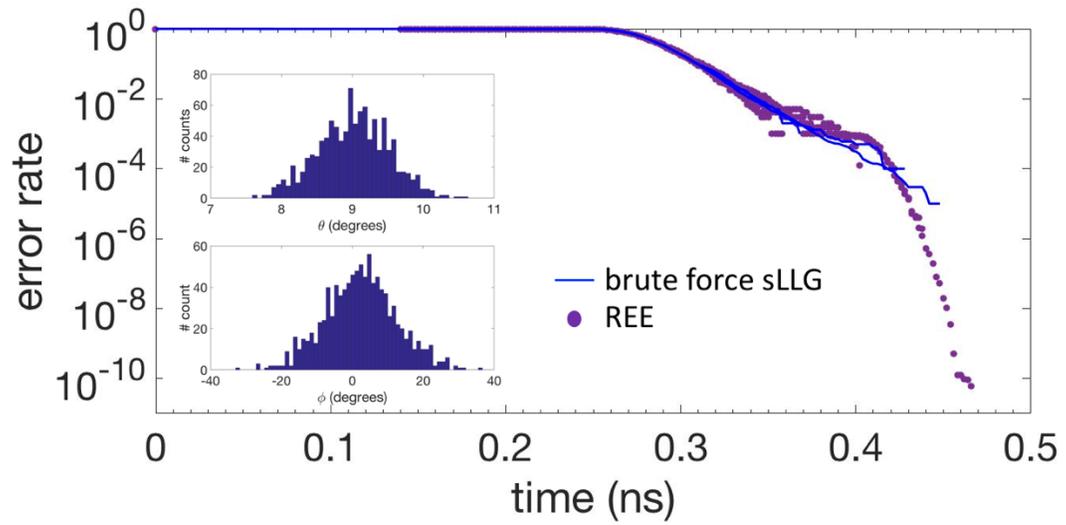

Figure 8. Approximate error rate estimation using equivalent single domain magnetization and performing brute force sLLG simulation and REE technique.